# Foundation Models in Biomedical Imaging: Turning Hype into Reality

Amgad Muneer[1], Kai Zhang[1], Ibraheem Hamdi[2], Rizwan Qureshi [3,4], Muhammad Waqas[1],

Shereen Fouad[5], Hazrat Ali[6], Syed Muhammad Anwar [7,8], Jia Wu[1,9]✉

1. Department of Imaging Physics, The University of Texas MD Anderson Cancer Center, Houston, TX 77030, USA
2. Center for Secure Artificial Intelligence for Healthcare (SAFE), McWilliams School of Biomedical Informatics, UTHealth Houston, TX, USA
3. Female Medicine in Machine Learning, Massachusetts Institute of Technology, MA, USA
4. Department of Computer Science, Salim Habib University, Karachi, Pakistan
5. School of Computer Science and Digital Technologies, Aston Centre for Artificial Intelligence Research and Application, Aston University, UK
6. Division of Computing Science and Mathematics, University of Stirling, Stirling, UK
7. School of Medicine and Health Sciences, George Washington University, Washington, DC, 20052, USA
8. Sheikh Zayed Institute for Pediatric Surgical Innovation, Children's National Hospital, Washington, DC, 20010, USA
9. Department of Thoracic/Head and Neck Medical Oncology, The University of Texas MD Anderson Cancer Center, Houston, TX 77030, USA

**Corresponding Author**

Jia Wu, PhD

Department of Imaging Physics

Department of Thoracic/Head and Neck Medical Oncology

The University of Texas MD Anderson Cancer Center

1515 Holcombe Blvd

Houston, TX 77030, USA

Telephone: 713-563-2719

e-mail: jwu11@mdanderson.org

**ABSTRACT**

Foundation models (FMs) are driving a prominent shift in biomedical imaging from task-specific models to unified backbone models for diverse tasks. This opens an avenue to integrate imaging, pathology, clinical records, and genomics data into a composite system. However, this vision contrasts sharply with modern medicine's trajectory toward more granular sub-specialization. This tension, coupled with data scarcity, domain heterogeneity, and limited interpretability, creates a gap between benchmark success and real-world clinical value. We argue that FM's immediate role lies in augmenting, not replacing, clinical expertise. To separate hype from reality, we introduce REAL-FM (Real-world Evaluation and Assessment of Foundation Models), a multi-dimensional framework assessing data, technical readiness, clinical value, workflow integration, and responsible AI. Using REAL-FM, we find that while FMs excel in pattern recognition, they fall short on causal reasoning, domain robustness, and safety. Clinical translation is hindered by scarce representative data for model training, unverified generalization beyond over-simplified benchmark settings, and a lack of prospective outcome-based validation. We further examine FM reasoning paradigm, including sequential logic, spatial understanding, and symbolic domain knowledge. We envision that path forward lies not in a monolithic medical oracle, but in coordinated subspecialist AI systems that are transparent, safe, and clinically grounded.

## 1. Introduction

The development and deployment of artificial intelligence (AI) is undergoing a paradigm shift, moving from an era of models trained on relatively small datasets for specific tasks to a dominant foundation model (FM) capable of multiple tasks. In this pursuit, there are two fundamental innovations in this new paradigm: 1) model training on big data with an aim to perform well in a wide variety of downstream tasks; 2) model pre-training is usually self-supervised and does not rely on the available ground truth labels. Under scaling law, billion-parameter models, pretrained on vast and diverse datasets, have demonstrated cross-task generalization in a multitude of downstream applications with remarkable flexibility.[1,2] This paradigm shift also holds profound implications for biomedical engineering in general and medical imaging in particular, a field defined by its complex, multimodal high-dimensional data, where getting high-quality ground truth labels at scale is a big challenge[3]. Specialized AI methods have already demonstrated success in discrete tasks like identifying diabetic retinopathy[4], segmenting tumors[5] or detecting breast cancer[6], and others[7] using medical images. However, FMs promise a more holistic, generalist intelligence capable of integrating diverse data sources (both structured and unstructured) such as imaging, electronic health records, genomics, and clinical notes to emulate the comprehensive reasoning of a human expert.[3] While the scaling law of general-domain AI suggests a trajectory toward a single, monolithic medical oracle, the heterogeneous and dynamic nature of clinical data and the long tail of rare pathologies suggest that clinical reality lies in modular generalist platforms, which are ecosystems of specialized experts coordinated by a central orchestrator.

This potential of current AI technology has generated immense excitement, and herald a future of augmented diagnostics, personalized treatment planning, precision medicine, and streamlined clinical workflows.[8–10] However, it has also fueled considerable hype, and created a critical need to distinguish between their future potential from current limitations. Two major challenges that hinder the current FMs strategy for real-world medical data are: 1) The very data that fuels generalist FMs is a primary obstacle in healthcare; the scarcity of large-scale, diverse, and representative medical datasets is compounded by strict patient privacy concerns, further complicated by regulatory frameworks, creating a stark contrast with the web-scale curated data used to train general-domain FMs.[11–14] 2) The opacity of deep learning architectures, often referred to as the black box problem, precludes the auditability required for high-stakes clinical accountability (trust[15], reliability[16], and accountability[17]). An incorrect algorithmic diagnosis often carries high risks. In addition, the narrowness of AI models is another persistent issue: a model that excels in one imaging modality or clinical context may fail when deployed in another, highlights the difficulty of assessing generalization in the heterogeneous landscape of healthcare.[18] Quantifying these technical and ethical hurdles is essential to move beyond benchmark-centric evaluation toward reliable clinical utility metrics.

We hypothesize that the immediate utility of FMs lies in assistive diagnostic augmentation rather than autonomous clinical reasoning. Current architectural paradigms must transcend benchmark leaderboard success to incorporate genuine clinical reasoning. We explore the frontiers of implicit, spatial, and symbolic reasoning, and examine how current models attempt to emulate the cognitive processes of domain experts. We further examine the current state-of-the-art by navigating the hurdles of data scarcity, validation, ethical deployment, and logistics to construct a comprehensive and discerning overview. Our analysis shows that while the vision of a fully autonomous AI-medical generalist remains on the horizon, the immediate reality is the emergence of a powerful new class of assistive tools for diagnosis and precision medicine. These models/tools are beginning to augment, rather than replace human expertise, and understanding their true capabilities and limitations is the first step toward responsibly integrating them into the future of clinics. By benchmarking current SOTA against clinical requirements, we demonstrate that while FMs excel at pattern recognition, they lack the causal-aware reasoning required for independent practice. We think the evolution of AI into three distinct phases: the *Hype* era of foundational architectures, the current *Reality Check* where deployment hurdles predominate, and a *Future* defined by neuro-symbolic and causally-aware systems, see **Fig. 1**.

### 1.1 REAL-FM: A Framework for the Real-world Evaluation and Assessment of Foundation Models

To bridge the gap between benchmark performance and clinical translation, we introduce **REAL-FM** (**R**eal-world **E**valuation and **A**ssessment of Foundation Mode**L**s for Biomedical Imaging), the first structured framework, to our knowledge, designed specifically to evaluate imaging FMs against the rigorous requirements of clinical practice. REAL-FM was developed through an international collaboration among computer scientists, and translational AI researchers, informed by a systematic analysis of the recent literature on biomedical FMs and established guidelines for clinical AI evaluation. Unlike existing benchmarks that privilege technical metrics such as accuracy or AUC, REAL-FM integrates five critical dimensions, data, technical readiness, clinical value, workflow integration, and responsible AI (**Table 1**), each operationalized with paired *indicators of hype* and *indicators of reality*. The framework's purpose is threefold: (i) to provide researchers with a diagnostic checklist for self-assessing FM claims before publication or clinical translation, (ii) to give clinicians and healthcare systems a transparent basis for evaluating vendor claims, and (iii) to identify which axes currently lag in the field, thereby directing future research toward the most translationally consequential gaps. By applying REAL-FM, we identify that technologies excelling in isolated performance metrics, yet lacking in workflow integration or responsible AI, are unlikely to deliver sustainable clinical value.

We do not merely catalog recent models; rather, we use this framework as an evaluative lens to critically assess current applications (Section 2), deconstruct reasoning paradigms (Section 3), and define the technical and regulatory requirements for clinical maturity (Sections 4 and 5). **Table 1** provides a checklist of *indicators of hype and reality* that any researcher or stakeholder can use to conduct a structured evaluation of an FM-based research claim. To assist readers across disciplines, **Box 1** provides definitions of key technical terms used throughout this Perspective.

**Box 1**. Glossary of Key Terms

| Term | Technical Definition |
|---|---|
| **Agentic AI** | Agentic AI refers to autonomous systems that plan and execute tasks to achieve high-level goals through multi-step workflows. |
| **Causality** | Causality focuses on cause-and-effect relationships to distinguish true drivers of change from incidental spurious correlations. |
| **Chain-of-Thought** | Chain-of-thought prompting forces a model to generate intermediate steps to improve the accuracy of complex logical deductions. |
| **Clinical Readiness** | Clinical readiness measures whether a system meets the safety and integration benchmarks required for live medical workflows. |
| **Contrastive Learning** | Contrastive learning is a training strategy that minimizes the distance between similar data pairs within an embedding space. |
| **Domain Shift** | Domain shift refers to a discrepancy between training and operational data that often reduces model reliability in new environments. |
| **Embedding** | A learned numerical representation (vector) that captures the semantic content of an input (image, text, or data point) in a compact form. |
| **Federated Learning** | Federated learning is a decentralized training technique that allows models to learn from local data sources without data transfer. |
| **Foundation Model** | A foundation model undergoes pretraining on broad datasets and serves as a base for multiple downstream applications. |
| **Grounding** | Grounding links model outputs to verifiable data, such as mapping text labels to specific coordinates in an image. |
| **Hallucination** | A hallucination is a model error where the output is factually false despite remaining grammatically correct. |
| **Knowledge Graph** | A knowledge graph provides a structured network of entities and relationships to serve as a machine-readable framework for expertise. |

| Term | Technical Definition |
|---|---|
| **Mixture of Experts** | A mixture of experts' architecture uses specialized sub-networks to increase model capacity without a linear increase in compute costs. |
| **Neuro-Symbolic AI** | Neuro-symbolic AI combines the pattern recognition of neural networks with the formal logic of symbolic systems. |
| **Symbolic Reasoning** | Symbolic reasoning uses explicit rules and logic to ensure a model follows strict constraints rather than statistical probability. |
| **Uncertainty Quantification** | Uncertainty quantification consists of methods that calculate prediction confidence levels to identify outputs requiring human intervention. |
| **Vision-Language Model** | A model jointly trained on images and text to learn aligned representations across both modalities. |
| **World Model** | An internal representation that enables an AI system to simulate and predict outcomes within an environment, allowing imaging FMs to reason about anatomy, disease progression, and treatment response beyond statistical pattern matching. |
| **Zero-shot Learning** | Zero-shot learning enables a model to complete tasks without specific training examples by leveraging general knowledge. |

## 2. Assessing Current Applications through the Multi-Axis Framework REAL-FM

Clinical FMs diverge from traditional AI by decoupling functional applications (output-centric tasks like segmentation and visual question answering, as outlined in Section 2) from reasoning paradigms (internal logic such as spatial or symbolic processing, as outlined in Section 3). This bifurcation is evidenced by models reaching technical readiness in task execution while simultaneously failing responsible AI benchmarks due to opaque or unverifiable reasoning (**Table 1**).

The rise of FMs signals a profound shift in how AI influences medical practice, and move beyond traditional pattern recognition to the simulation of higher-level cognitive reasoning and decision-making processes.[19] For years, medical AI has been dominated by task-specific models, developed to perform narrow functions like classifying a specific pathology or segmenting a particular organ or lesion.[18] These models, while often effective, function as digital specialists, with their thinking and grounding confined to predefined rules and labeled examples (the ground truth) provided to guide the model training.[20] FMs, in contrast, have introduced the prospect of unified systems capable of contextual reasoning and contextual comprehension across a spectrum of downstream tasks, approaching the complex, multifactorial decision-making of human clinicians.[19]

However, whether FMs truly reason or merely reproduce statistical patterns remains unresolved. Clinical reasoning requires integrating multimodal data across sequential and parallel investigations, where each finding refines a developing hypothesis.[3] In this section, we examine the clinical applications of FMs and evaluate their underlying reasoning paradigms against the REAL-FM framework established in **Table 1**.

### 2.1 Medical Image Analysis

FMs facilitate zero-shot and weakly supervised generalization across segmentation, classification, and prognostic tasks.

#### 2.1.1 Segmentation

Recent advances in FMs for medical image segmentation have been marked by the emergence of universal, multimodal architectures capable of zero-shot and weakly supervised learning across diverse imaging modalities. Early advances were led by models like MedSAM[5], which adapted the Segment Anything Model (SAM) to the medical domain by retraining on over a million annotated images spanning diverse modalities such as CT, MRI, X-ray, and ultrasound.[21] MedSAM demonstrated that a single architecture could provide promotable, interactive segmentation masks for a variety of anatomical structures without requiring organ or modality-specific retraining. However, it still showed limitations in segmenting fine structures and required well-designed prompts for reliable results.[5] SAM-Med2D extended this paradigm by fine-tuning SAM on an even larger and more heterogeneous medical dataset, freezing the core encoder, and introducing adapters to boost medical robustness.[22] Building on these 2D advancements, SAM-Med3D introduced a fully learnable 3D architecture and enables accurate volumetric segmentation across diverse modalities and anatomical structures using sparse 3D prompt points.[23] While early medical FMs struggled with fine structures, recent architectures like MedSegX[24] have begun to satisfy the *Reality* criteria of REAL-FM framework. MedSegX introduces a Contextual Mixture of Adapter Experts within a generalist architecture trained on MedSegDB, a publicly accessible database spanning 39 organs and 10 modalities, and achieve state-of-the-art performance in both in-distribution and out-of-distribution settings, including zero-shot generalization to unseen clinical cohorts.[24] BioMedCLIP took a different approach by training a vision-language model (VLM) on millions of biomedical images and paired text, enabling not just visual but also semantic understanding of medical content.[25] MedCLIP-SAMv2 further bridges vision and language by coupling BiomedCLIP with SAM to generate segmentation masks directly from text prompts[26], outperforming prior zero-shot methods (SaLIP[27], SAMAug[28], MedCLIP-SAM[29]) while narrowing the gap to fully supervised models like nnUNet.[30]

While these models outperform earlier iterations (SaLIP, SAMAug), a performance gap remains, fully supervised models (e.g., nnU-Net) consistently yield higher Dice scores, a standard measure of segmentation overlap accuracy, on benchmark modalities, including breast ultrasound and brain MRI (**Suppl. Fig. 1,** adapted from the literature). FMs are also reshaping cellular and microscopy imaging, where accurate segmentation underpins spatial transcriptomics, live-cell phenotyping, and high-content drug screening. Traditional tools have been constrained to specific modalities or organism classes, necessitating retraining for each new condition. Segment Anything for Microscopy (uSAM) overcomes this by fine-tuning generalist SAM models for light and electron microscopy, achieving robust segmentation across fluorescence, phase-contrast, and electron modalities.[31] CellSAM takes a complementary approach, coupling a dedicated cell detector (CellFinder) with SAM's mask generator to achieve human-level performance across mammalian cells, yeast, and bacteria with strong zero-shot transfer.[32] Under our framework REAL-FM, these cellular FMs demonstrate high technical readiness through cross-modality generalization but lack clinical value validation through prospective deployment and responsible AI assessment of failure modes across rare cell morphologies.

### 2.1.2 Classification

FMs are fundamentally changing how pathologies are classified by extracting rich, generalized feature representations from pre-training. Unlike traditional deep learning models that require extensive labeled data for each specific disease, FMs enable transfer learning and few-shot classification by fine-tuning a pre-trained general medical representation. Models trained with Contrastive Learning on massive image-text pairs (such as a generic VLM adapted for the medical domain) learn embeddings where an image's visual features are closely aligned with its corresponding textual report (e.g., malignant nodule)[25]. This capability allows for highly accurate zero-shot classification of pathologies, like classifying chest X-rays for pneumonia, pneumothorax, or cardiomegaly[33], even when the model has never explicitly been trained on labeled images for all three conditions. The clinical reality, however, is that performance often degrades significantly on external, out-of-distribution datasets, underscoring the gap between benchmark classification accuracy and robust, real-world deployment[34].

### 2.1.3 Detection and Localization

The task of detection, identifying and localizing specific clinical entities, such as small lesions, nodules, or fractures, benefits significantly from the global context provided by FMs. FMs can develop an implicit understanding of normal anatomy by leveraging self-supervised training on a vast corpus of images[35]. This foundational knowledge allows FMs to more effectively spot and delineate abnormalities (out-of-distribution regions) that deviate from the learned normal distribution[36]. Techniques like masked image modeling, a common pre-training strategy, force the model to predict hidden parts of the image, generating superior internal representations that improve the localization of subtle findings[37]. For instance, in mammography or CT scans, FMs are adapted to serve as powerful CAD (Computer-Aided Detection) systems that highlight potential areas of concern, thereby reducing the rate of missed diagnoses. A persistent challenge is managing the trade-off between sensitivity and specificity in high-stakes diagnostic settings.

### 2.1.4 Prognosis and Prediction

FMs are increasingly being applied to prognosis and outcome prediction, tasks that require integrating longitudinal data and modeling disease trajectories rather than classifying static snapshots. This involves utilizing the model's holistic, multi-modal feature integration capabilities to forecast future patient states, such as the likelihood of disease progression, response to a specific therapy, or risk of a future adverse event (e.g., heart attack, stroke)[36]. FMs can generate powerful, personalized risk scores by integrating imaging biomarkers with data from Electronic Health Records (EHRs), including demographics, lab values, and clinical history[38]. The prediction task demands a deeper understanding of causal inference rather than mere statistical correlation, as predicting why a patient might decline is more clinically valuable than simply predicting that they will[39]. The path to clinical reality for predictive models is the most challenging, as it requires rigorous temporal validation on prospective data to ensure their predictions hold across long-term follow-up and diverse patient populations.

### 2.2 Biomedical Visual Question Answering and Grounding

Biomedical Visual Question Answering (VQA) aims to enable AI systems to answer clinical questions about medical images by integrating visual and textual reasoning. Early models relied on discriminative approaches trained on curated datasets like VQA-RAD[40] and PathVQA[41]. The field has since advanced toward large-scale multimodal pretraining, as seen in models like BioViL[42], BiomedCLIP[25], and UniBiomed[43], which leverage millions of image–report pairs to learn joint representations. More recently, biomedical adaptations of generalist VLMs such as Med-PaLM[38], LLaVA-Med[44], GatorTron-MoL[45], Med-Gemma[46], PathChat[47] and the recently published RadFM[48] have enabled open-ended, instruction-tuned VQA grounded in a clinical context. These systems combine FMs' scale with domain-specific supervision, and move closer to interactive, explainable image understanding. Despite these gains, biomedical VQA still faces key challenges: generalizing across pathologies, handling rare conditions, hallucinations[49] , grounding answers in image evidence, and providing reliable uncertainty estimates. Benchmarking remains limited, often focusing on answer accuracy without capturing clinical relevance or safety.

Grounding refers to the model's ability to localize and interpret image regions when answering questions or generating annotations, which is critical for clinical trust and regulatory compliance.[50] Recent efforts like MIMO (Medical Vision–Language model with Visual Referring and Pixel

Grounding)[51] supports both visual prompts and segmentation-based grounded outputs by aligning LLM-generated tokens with segmentation masks. UniBiomed merges an LLM with a SAM-like decoder to deliver unified outputs: clinical text explanations with corresponding segmented regions across modalities. Benchmarks like HEAL-MedVQA[50] and Localize-before-Answer (LobA)[50] show that explicit segmentation-based grounding meaningfully reduces hallucinations and improves accuracy. To this end, grounding is shifting from a theoretical ideal to an operational tool. By combining segmentation masks and visual prompts, modern systems can focus their reasoning on clinically relevant areas and produce verifiable outputs, paving the way for systems that explain what they see, where they see, and why this is essential for clinical translation.

Despite meeting technical readiness indicators of REAL-FM, current VQA models exhibit high hallucination rates and lack standardized safety benchmarking. The transition from discriminative to generative grounding (exemplified by MIMO and UniBiomed) shifts these systems from theoretical prototypes to operational tools. However, the extent to which these outputs reflect clinical reasoning versus surface-level pattern matching remains a critical barrier to clinical translation.

### 2.3 REAL-FM-Based Evaluation of Translational Readiness

To systematically assess the translational readiness of biomedical foundation models, we evaluated 15 representative FMs across four domains: segmentation, classification and retrieval, pathology, and generalist multi-task systems. Each model was assessed against the reality criteria (**Table 1**). Two reviewers (A.M. and K.Z.) independently assigned a score from 0 to 10 for each evaluation dimension and provided criterion-specific justifications for every rating. Any scoring disagreements were adjudicated by a third reviewer (R.Q.). Although several models demonstrated strong performance on benchmark-oriented dimensions, substantial deficiencies persisted in clinically critical areas including calibration, usability, interoperability, privacy, accountability, and bias and equity assessment. These findings highlight a persistent gap between technical progress and real-world clinical translation (**Fig. 2**).

## 3. Reasoning Paradigms: Distinguishing Cognitive Reality from Statistical Mimicry

While Section 2 established that FMs can execute medical tasks, clinical reality of REAL-FM demands that these tasks are underpinned by robust logic. We therefore treat reasoning as a distinct evaluative category. An FM without explicit reasoning remains a black box. It satisfies technical benchmarks through statistical correlation while failing the safety and trustworthiness requirements that are essential for high-stakes medical intervention.

Evaluating FMs' reasoning capabilities must be contextualized within the inherent limitations of human cognition. Even expert radiologists are susceptible to biases such as anchoring on initial impressions, prematurely ceasing their search after detecting one abnormality, or aligning with prevailing diagnostic opinions. Recognizing these tendencies emphasizes that both human and machine reasoning warrant scrutiny through a common lens of interpretative fallibility[52]. These mental shortcuts and shortcut learning, or heuristics, highlight that the human benchmark is not one of perfect, infallible logic but of a complex, experience-driven, and sometimes flawed process. As we will see later sections, FMs exhibit their own analogous, and sometimes identical, failure modes.[53]

To deconstruct the capabilities and limitations of reasoning in biomedical FMs, this section provides a structured analysis across three key paradigms: 1) implicit and sequential reasoning, which emulates the step-by-step diagnostic workflow; 2) spatial reasoning, which involves understanding complex anatomical and pathological context; and 3) explicit and symbolic reasoning, which aims to integrate verifiable domain knowledge. This taxonomy, summarized in **Table 2**, provides a framework for critically assessing whether FMs' reasoning capabilities are a tangible reality or persistent hype.

### 3.1 Implicit and Sequential Reasoning

The first major frontier for reasoning in FMs is the emulation of the sequential, step-by-step process that characterizes clinical investigation. This capability has been unlocked mainly by a prompt engineering technique known as Chain-of-Thought (CoT)[54]. CoT guides a model to break down a complex problem into a series of intermediate, manageable steps by simulating a human-like reasoning process. The application of CoT has shown significant promise in the medical field.[55] Early studies demonstrated that CoT prompting could enhance the performance of even smaller language models on medical question-answering tasks. This principle has been integrated into state-of-the-art medical LLMs, where medical VQA datasets such as MS-ROCO[56], VQA-RAD[40] makes structured reasoning more accessible.[57] For instance, Med-PaLM-2 explicitly incorporates advanced strategies such as ensemble refinement and a chain of retrieval to enhance its reasoning capabilities, contributing to its high performance on medical licensing exam (USMLE) questions.[57] These models can produce detailed, step-by-step diagnostic evaluations that emulate the differential diagnosis process of a clinician, systematically considering and ruling out possibilities based on the provided case information.

However, for biomedical imaging, purely textual reasoning is insufficient. The critical evolution has been the development of multimodal frameworks that attempt to ground these textual reasoning chains in visual evidence from the images themselves. This represents a significant step towards verifiability. One such framework, MedCoT[58], addresses the inherent unreliability of a single AI expert by creating a hierarchical verification chain. In this system, an initial specialist model first proposes a diagnostic rationale based on the image and text query. A follow-up specialist model then reviews this rationale, either validating it or forcing a reassessment. Finally, a diagnostic specialist model, implemented as a sparse Mixture-of-Experts (MoE), votes on the refined rationale to produce a final, consensus-based answer. The Chain-of-Visual-Thought (CoVT) framework takes this visual grounding a step further.[59] Recognizing that a textual rationale alone can be untethered from the visual facts, CoVT aims to create a more interpretable and accurate diagnostic process (***Suppl Fig. 2***) by explicitly linking individual statements in a generated radiology report to corresponding visual prompts within the image, such as masks, landmarks, or bounding boxes, such as CoVT-CXR[59] . The development of these sophisticated frameworks leaves a lasting impact, as the field moves towards powerful, verifiable reasoning models. This shift from explainable AI, where the model simply explains its actions, to trustworthy reasoning AI, where it must provide evidence that can be checked, is much more demanding in terms of data and engineering than the hype around autonomous reasoning makes it seem. This engineered scaffolding is necessary because, despite these advances, FMs' reasoning remains brittle. The models are still prone to generating plausible but misleading or entirely fabricated reasoning chains.[60] Detailed error analyses of advanced LLMs like DeepSeek R1 reveal a striking parallel to human cognitive failures; the models exhibit clear instances of *anchoring bias*, *difficulty integrating conflicting data*, and a *limited consideration of alternative diagnoses*.[53] Several studies have found that longer, more detailed reasoning chains from these models are often correlated with a *higher* probability of error (**Fig. 3a)**.[61] This surprising finding suggests that, in many cases, overthinking kills[62]; long explanations are not a sign of deep reasoning but rather a way for the model to justify a wrong answer it has already reached through shallow statistical patterns (**Fig. 3b & 3c**).

### 3.2 Spatial Reasoning: Interpreting Complex Anatomical and Pathological Contexts

Beyond sequential logic, clinical diagnosis in imaging-centric fields such as radiology, pathology, and ophthalmology depends on *spatially grounded interpretation*: understanding 3D anatomy, localizing abnormalities, relating findings to nearby critical structures, and integrating context across views and scales. For example, when interpreting a brain MRI, a radiologist synthesizes axial, sagittal, and coronal planes to localize a lesion, determine its relationship to eloquent cortex and vascular territories, assess mass effect, and judge whether changes are clinically meaningful over time. These capabilities extend beyond isolated 2D pattern recognition.[63] Accordingly, while 2D analysis has been central to medical AI, many clinically relevant endpoints require robust reasoning in 3D, because volumetric CT/MRI encode the spatial context needed for accurate measurement,[64] procedural planning,[65] and longitudinal monitoring.[66] In this settings, effective interpretations depends on an implicit world model, as an in-silico representation of anatomical structure, spatial relationships, and physic laws that enables consistent reasoning across views, scales, and time[67]. In practice, segmentation and localization are not reasoning by themselves, but they provide the *spatial grounding* that higher-level reasoning depends on defining *where* a structure is, *what* it is, and *how* it relates to surrounding anatomy.

A key challenge for scalable spatial intelligence is moving beyond treating 3D volumes as stacks of 2D slices and instead modeling their intrinsic geometry and correspondence across scans.[68] In medical imaging, the geometric correspondence problem refers to identifying matched anatomical structures across different images despite variability in orientation, pose, acquisition protocol, and patient-specific anatomy.[69] This is foundational for precise localization, treatment planning, and image-guided interventions. Geometric Deep Learning (GDL) addresses this by operating on non-Euclidean representations (graphs, meshes, or point clouds) where anatomy can be represented compactly and manipulated in anatomically meaningful coordinates.[70] For example, representing an organ surface as a point cloud can substantially reduce data size compared with dense voxel grids while preserving the geometry needed for registration, tracking, and measurement.

Within this broader view, progress in spatial reasoning is increasingly driven by modular systems that combine spatial grounding, geometric representation, and task-specific decision layers, rather than by a single monolithic generalist. The evolution from limited transfer of generalist segmenters such as SAM, to domain-adapted variants such as MedSAM[5], to architecturally adaptive systems (e.g., SPAD-Nets[71]) and federations of experts (e.g., MoME[72]) supports a consistent conclusion: the near-term future is more plausibly a generalist platform rather than a single generalist model[73], where specialized components are invoked depending on modality, anatomy, and clinical question. This trend is especially pronounced in computational pathology and radiology (**Fig. 4**), where FMs are increasingly designed to capture complex structural and morphological hallmarks from specific imaging modalities.[74,75] In histopathology, vision-only FMs such as UNI[75], PLUTO[76], GPFM[77], PathChat[47] and RudolfV[78], trained on millions of whole-slide images (WSIs), serve as general-purpose morphology encoders adaptable to tasks from cancer classification to cell segmentation. Multimodal pathology FMs such as PRISM[79], CONCH[80], TITAN[81], CPath-Omni[82] and MUSK[83] go further by aligning image features with clinical text to enrich contextual grounding. MUSK[83], trained on 50 million pathology images and one billion text tokens via unified masked modeling, achieves superior performance across 23 benchmarks, including biomarker prediction, visual question answering, and immunotherapy response prediction, without task-specific fine-tuning, demonstrating that multimodal pathology FMs are increasingly meeting the technical readiness criteria outlined in **REAL-FM**.

Parallel efforts in radiology have produced dedicated models for high-volume modalities such as CT. CT-CLIP, for example, is trained on paired 3D chest CT volumes and radiology reports[84], strengthening image–text correspondence and contextual grounding. Similarly, Google's CT-FM is a large-scale 3D FM pretrained on over 148,000 CT scans to support diverse downstream radiological tasks.[85] Collectively, these models do not automatically confer reasoning, but they substantially improve the *spatial and semantic grounding* on top of which clinically meaningful reasoning can be built. Notably, several pathology FMs are beginning to satisfy the "Reality" indicators in **REAL-FM**; TITAN[81] and MUSK[83] demonstrate multi-site validation, clinically relevant endpoints (survival prediction, treatment response), and strong out-of-distribution generalization, suggesting that computational pathology is among the first imaging domains where FMs are starting to transition from hype to operational utility. However, gaps remain on other axes, including the absence of prospective validation (clinical value), deployment in clinical workflows (workflow integration), and systematic evaluation of fairness and uncertainty (responsible AI), emphasizing that even the most advanced pathology FMs have not yet fully crossed the hype-to-reality threshold. This reinforces that the path toward robust spatial reasoning in medicine is likely to emerge from a diverse ecosystem of domain-specialized components, integrated into modular platforms with explicit objectives and evaluations for spatially grounded clinical decision-making.

Despite the impressive breadth of these specialized vision FMs, the field faces a fundamental architectural tension: the *One Model to Rule Them All* narrative, i.e., a cornerstone of general-domain scaling laws, encounters a data scarcity wall in the clinical environment. In the pursuit of a monolithic medical oracle, large-scale models often suffer from *representation collapse,* where the statistical dominance of common pathologies (e.g., pleural effusion or fatty liver) washes out the subtle, high-frequency features required to identify the long tail of rare diseases, such as specific pediatric osteosarcomas or orphan genetic disorders. Extreme scaling on diverse datasets can inadvertently degrade performance on niche, high-stakes diagnostic tasks.

Consequently, the operational reality of FMs in imaging is shifting toward Modular Generalist Platforms. In this paradigm, a foundational reasoning backbone, acting as a cognitive router, does not attempt to internalize every anatomical nuance. Instead, it leverages a mixture-of-experts architecture or an agentic framework to orchestrate a suite of specialized heads or sub-models. This approach ensures that the precision of domain-specific encoders is preserved while the generalist backbone provides the cross-modal synthesis and sequential logic necessary for complex clinical reasoning.

### 3.3 Explicit and Symbolic Reasoning: Integrating Domain Knowledge for Robustness

While sequential and spatial reasoning represent significant advances, they are largely based on learning statistical correlations from data.[86] Purely data-driven models can learn spurious correlations, are prone to instability when faced with dataset shifts, and lack the deep interpretability required for clinical trust.[87] This has led to a renewed focus on explicit and symbolic reasoning, a paradigm that seeks to integrate structured, human-curated domain knowledge directly into the FMs pipeline.[88] The primary vehicle for this integration is the medical Multimodal Knowledge Graph (MKG)[89,90], a structured representation of entities (e.g., lung, nodule, malignancy) and their relationships (e.g., "is located in", "is a type of"). MKGs provide a formal, machine-readable embodiment of declarative (what is known) and procedural (how-to) knowledge that can be used to guide and constrain the learning process of a neural network. For example, Knowledge-Boosting Contrastive Vision-Language Pre-training (KoBo) framework[91], identified critical failure modes in standard vision-language pre-training for radiology, such as the semantic overlap problem (where different diseases like pneumonia and edema can have similar visual features) and the semantic shifting problem (where a single term like opacity can refer to vastly different visual findings). To combat this, KoBo integrates a clinical MKG directly into the contrastive learning objective.[91]

Beyond training, symbolic representations, including knowledge graphs, rule-based systems, and ontologies, are also proving essential for more rigorous and clinically relevant evaluation. Standard text-similarity metrics for tasks like radiology report generation, such as BLEU or ROUGE, are notoriously poor proxies for clinical accuracy, as a report can be lexically similar to a reference yet diagnostically incorrect.[92] Symbolic methods instead evaluate predictions at the level of structured clinical facts and rules: for example, assessing whether the generated report correctly encodes entities (e.g., lesions, organs, devices), their relations (e.g., “metastatic to,” “invading”), and their consistency with guideline-derived rules (e.g., staging criteria, RECIST definitions). Within this family, the Radiology Knowledge Graph (RxKG) system moves evaluation from the text level to the knowledge level by comparing model outputs against a curated graph of radiologic findings and relations.[93] More generally, ontologies such as RadLex or SNOMED CT, coupled with symbolic reasoners, can be used to detect contradictions, missing findings, or violations of clinical guidelines in model outputs[94]. Human-level evaluation remains critical, as expert radiologists still catch subtle errors and hallucinations missed by automated metrics. Yet, symbolic and KG-based evaluations provide a scalable, clinically aligned layer of oversight that is far better matched to the goals of FMs than surface-level text scores.[95]

Current evidence paints a dual picture: FMs demonstrate strong reasoning performance on structured benchmarks but remain brittle when evaluated at deeper, knowledge-grounded levels. A synthesis of the current state-of-the-art, summarized in **Table 3**, illustrates this duality. Models like Med-PaLM 2 have achieved expert-level performance on standardized medical exams, with clinicians preferring its long-form answers to those of generalist physicians across multiple axes, including reasoning.[57] This demonstrates a profound capability to process and synthesize medical knowledge. However, the crucial caveats are that specialist physicians' answers are still preferred overall, and the model's safety is rated as only on par with, not superior to, human physicians.[57] Similarly, generalist vision-language models like MedVersa[96], PathChat[47], Med-CLIP[97], and RadFM[48] show strong, sometimes state-of-the-art, performance across a range of medical tasks, with generated reports that can match or exceed human reports in a majority of cases. Yet, at a deeper, knowledge-based level, evaluations using frameworks like ReXKG reveal that these models still struggle to capture the full context of relationships and concepts in expert-human reports, indicating a gap in comprehensive understanding.[93]

To sum up, FM reasoning is real but conditional: it depends on domain-specific data, deliberate architectural design, and continuous human oversight. Evaluated against the REAL-FM framework in **Table 1**, current models increasingly satisfy technical readiness criteria but fall short on clinical value and responsible AI axes, specifically regarding prospective validation, uncertainty quantification, and verifiable safety. Closing these gaps is the focus of the sections that follow.

## 4. The way forward: from hype to reality

The development of FMs is catalyzing a conceptual leap in clinical AI, and shift the paradigm from passive, task-specific tools to autonomous agents capable of independently executing complex clinical workflows (**Fig. 5**).[98] This agentic shift is powerfully demonstrated by recent work, such as Ferber et al.'s autonomous oncology agent[99], which used an LLM not as a static knowledge base, but as a cognitive orchestrator for a suite of specialized, high-performance tools. It identifies that a case requires specific 3D segmentation of the pelvic floor, calls a specialized MedSAM-variant, retrieves relevant prior reports via RAG (Retrieval-Augmented Generation), and synthesizes these expert outputs into a causal reasoning chain. This router approach preserves the precision of specialized models while maintaining the flexibility of FMs. The integrated AI agent autonomously plans and executes complex workflows, ranging from analyzing histopathology slides via vision transformers to querying specialized knowledge bases. This systematic approach increased decision-making accuracy from 30.3% to 87.2%, a significant leap over the performance of GPT-4 alone.[99] This striking result reveals a critical principle: the agent's power derives not from knowing everything, but from its ability to find, synthesize, and reason upon information from expert sources, and function as an effective clinical teammate (***Suppl notes***).

Longitudinal imaging AI agents, which integrate temporal context, cross-modality segmentation, and guideline retrieval, render endpoint-only verification insufficient. Because errors may propagate through extended reasoning chains, safety requires a transition from output auditing to workflow-level explainability.[99] Clinicians must shift to active supervision, interrogating intermediate inference steps and re-anchoring AI logic within specific clinical constraints. This moves from passive verification to process-level oversight ensures human expertise remains the primary arbiter of diagnostic validity as system complexity increases.

Translating FMs into clinical practice requires progress on four fronts: causal understanding that moves beyond statistical correlation, inclusiveness across diverse populations and healthcare contexts, engineered trustworthiness with quantifiable uncertainty, and verifiable safety guarantees for high-stakes medical environments.

### 4.1 Causality: for higher-level intelligence

The next frontier for FMs in biomedical imaging is to transcend their current role as sophisticated pattern-matching engines and evolve into tools for causal inference (***Suppl Fig. 3***), thereby enabling genuine scientific discovery.[100] At present, even the most advanced models operate on probabilistic modeling that captures spurious correlations rather than true causal relationships.[101]

This limitation renders them “causal parrots”, capable of reciting learned associations from vast datasets without any underlying comprehension of biological mechanisms.[102] Causal Representation Learning (CRL) aims to uncover latent, high-level causal variables (e.g., disease processes, genetic expressions) from the observed, low-level data (e.g., image pixels).[103] This is achieved by leveraging multiple data distributions (arising from different scanners, patient populations, or time points) to identify invariant causal structures that persist across changing conditions. Furthermore, the use of Functional Causal Models (FCMs), such as linear non-Gaussian or post-nonlinear models, imposes structural constraints on the data-generating process[103], and make the causal direction identifiable from purely observational data. This is a critical advantage in medical imaging, where performing randomized interventions to establish causality is often difficult or unethical.[104] In biomedical imaging, CRL could learn to disentangle the causal factors of a tumor's appearance, separating the effects of its underlying biology from imaging artifacts or patient-specific anatomy. This capability is not merely an academic exercise; it is the key to solving persistent challenges, such as domain adaptation, where models must generalize across heterogeneous data sources.[105] The development of open-source libraries like causal-learn further promises to accelerate the adoption of these powerful methods.[106] To this end, integrating causality will transform FMs from diagnostic aids that answer *what* into instruments of scientific inquiry that can answer *why* and *how*. This opens the door to performing *silico* counterfactual experiments. For instance, asking a model *how would this glioblastoma's representation in a T1-weighted MRI change if we could hypothetically inhibit the EGFR pathway?* shifts the model's role from a passive pattern recognizer to an active, hypothesis-generation engine for automated scientific discovery.

### 4.2 Trustworthiness

Engineering clinical trust requires shifting from deterministic outputs to probabilistic calibration, where an FM's softmax response aligns with the empirical frequency of correct diagnoses across diverse patient cohorts.[107] The translation of FM raises complex trust issues that should be carefully considered throughout design, deployment, and validation. For example, privacy concerns arise when sensitive patient data are used; robustness issues occur due to possible variations in imaging modalities and patient populations; reliability issues also arise from the accurate versus inaccurate output, which is sometimes termed hallucination in generated reports; and fairness could be hampered if the model ends up reinforcing or furthering existing biases in healthcare data[108,109].

To move from high-level principles to deployable systems, the field can draw from focuses on probabilistic modeling and deep generative models, which, unlike deterministic models that provide a single, often overconfident, answer, can quantify their own uncertainty.[110] These models learn anatomical variations and, crucially, quantify aleatoric uncertainty (the inherent variability in data) which is essential for reliable predictions under challenging real-world imaging conditions.[11,84] The clinical significance of this approach is profound. A model that can express its own uncertainty, e.g., *I am 80% confident in this segmentation, but highly uncertain at the boundary with this critical structure*, provides actionable information for a clinician. It enables a robust human-in-the-loop workflow where high-uncertainty cases are automatically flagged for expert review, mitigating risk and building a synergistic human-AI partnership. This technical shift from seeking only accuracy to demanding calibrated uncertainty reframes trust from a subjective feeling into a quantifiable, verifiable property of the model. Regulatory bodies could approve an algorithm that is not based on a single accuracy metric, but on a tiered system in which predictions below a certain confidence threshold are mandated for human review.[111] This creates a certifiable, trustworthy human-AI system, and solve the governance problem that currently plagues medical AI.

In addition, augmenting FMs with explainability-informed tools is required to provide the rationale for FM decisions in medical imaging tasks. This includes attention maps or saliency overlays, supported by scores of uncertainty/confidence of prediction, textual explanation, and actionable information.[112] Privacy can be assured through federated learning or differential privacy,[108] the robustness can be achieved by multi-institutional heterogeneous dataset training and posing adversarial testing to uncover the potential vulnerabilities in the model.

### 4.3 Safety: From Benchmarking to Verifiable Guarantees

Ensuring patient safety requires a paradigm shift from the conventional performance-based evaluation of AI models to a rigorous safety engineering framework, and borrow principles of formal verification, risk-aware learning, and adversarial validation from other safety-critical domains.[113] In healthcare, safety is not a feature but a prerequisite.[114] The rapid integration of AI without sufficient governance, a recent survey found that only 16% of hospitals have system-wide AI governance policies, poses significant risks of flawed or biased decision-making[115]. Standard benchmarking on static datasets is dangerously insufficient for safety-critical deployment[115–117]. The safety paradigm moves beyond measuring average accuracy to include formal safety verification, which seeks to prove that a model adheres to specific, non-negotiable safety constraints. For a biomedical imaging model, such a constraint could be, e.g., *the segmentation boundary of a tumor must never encroach upon a predefined critical structure like the optic nerve*.

Falsification techniques would then be used to actively search for any input that could cause the model to violate this rule. This approach also demands multi-fidelity validation and adversarial robustness testing, where models are intentionally challenged with corrupted data and in realistic simulators to find failure modes before they can harm a patient.[118,119] This proactive search for vulnerabilities is especially critical given that over half of all AI-related failures stem from third-party tools[120]. It means researchers must shift focus from leaderboard-chasing to developing verifiably robust architecture. It requires regulators like the AHRQ and FDA to develop new certification frameworks based on formal verification and stress testing, beyond clinical trial performance data.[113] And it obligates hospitals to implement AI-Vigilance systems, analogous to pharmacovigilance, for continuous post-deployment monitoring and auditing. This promises a complete safety-first lifecycle for medical AI that is currently absent and transforms it from a promising but risky technology into a reliable and integral component of the healthcare infrastructure.

### 4.4 Inclusiveness

Inclusive FM design mitigates the current technical bias that frequently excludes healthcare workers' preferences and patients' values. Integrating inclusiveness across the model lifecycle, from problem formulation to post-deployment monitoring requires co-designing data and evaluation metrics with diverse stakeholders[92,121]. This multi-perspective approach ensures clinical relevance and seamless integration into biomedical imaging workflows, directly enhancing model trustworthiness, usability, and fairness[122,123]. By proactively obstructing bias, inclusive design stabilizes the clinician-patient relationship[124].

Beyond individual design, inclusiveness addresses systemic power imbalances between high-resource academic centers and under-resourced health systems in emerging economies. Bridging this gap requires formalized joint decision-making authority through multi-institution consortia and federated data partnerships. Such frameworks provide under-resourced partners with legislative and technical agencies over model updates and deployment protocols, ensuring global benefit-sharing rather than data extraction.

## 5. Clinical Translation and Integration

Technical maturity represents only the first phase of FMs development. The transition to clinical practice requires a distinct operational roadmap that addresses validation, regulation, and infrastructure. We outline four critical requirements for operationalizing FMs in clinical workflows.

### 5.1 Prospective Evidence and Outcome-Based Validation

Current shallow retrospective benchmarks rarely predict real-world utility. Clinical translation requires shifting the evaluation paradigm from in-silico metrics to prospective, randomized clinical trials. FMs introduce unique risks regarding automation bias and over generalization; therefore, validation must measure clinical outcomes, such as reduced time-to-diagnosis, improved morbidity, or altered treatment plans, clinical reasoning with real world grounding, rather than simple accuracy scores.[124] Furthermore, institutions should implement *silent trials*, where models run in the background without influencing care. This approach establishes performance baselines across heterogeneous local patient populations prior to active deployment.

### 5.2 Dynamic Regulation and Local Governance

Traditional regulatory frameworks, such as *Software as a Medical Device (SaMD)*, target static algorithms with deterministic outputs. FMs break this paradigm through their adaptability and open-ended generation. Regulation must therefore evolve from one-time clearance to continuous lifecycle monitoring (or AI-Vigilance, Section 4.3). This requires model cards that explicitly detail data provenance and known failure modes. Additionally, hospital governance boards must enforce local validation to ensure models trained on general biomedical corpora perform safely within specific institutional demographics, imaging protocols and emerging FMs as Medical Devices (FMaMD) regulations.

### 5.3 Workflow Integration and Interoperability

Integration friction remains a primary obstacle to adoption. FMs cannot function as standalone applications; they must embed directly into existing workflows via standards such as DICOM and HL7 FHIR. For radiology, this necessitates interfaces where FM outputs appear as preliminary layers within the native viewer. The user interface must support clinical reasoning by enabling clinicians to interrogate the model's process, for example by linking generated text to specific image regions or overlaying uncertainty maps on the original scan, rather than accepting opaque recommendations.

### 5.4 Economic Viability and Reimbursement

Widespread adoption depends on economic sustainability. The computational costs of FMs are significant and require healthcare systems to establish clear return on investment. This value derives from two sources: efficiency (e.g., automated volumetry increasing throughput) and quality (e.g., reduced liability from missed findings). Emerging FM capabilities including synthetic data generation, multi-modal integration of imaging with clinical records, further enhance scalability, and enable imaging AI systems to function as active diagnostic copilots rather than static decision-support tools. Reimbursement models must therefore evolve toward value-based frameworks that incentivize certified AI systems based on demonstrated clinical and operational impact and shift financial responsibility away from individual institutions. To move FMs from lab to bench, we need a transition from mono-modal, static, and opaque architectures to multi-modal, live, and transparent systems.

## 6. Perspective

The utility of imaging FMs now hinges on intended-use discipline, specifically, whether a model validated for screening-level sensitivity can maintain specificity when deployed in a high-prevalence diagnostic triage workflow. Clinical translation will therefore require standardized stress-testing protocols and prospective validation under realistic deployment conditions, especially in high-prevalence triage settings. Moving from hype to reality will require shifting the conversation from whether we can build large models to how they should be built, tested and used responsibly in biomedical imaging.

A practical starting point is intended-use discipline. Imaging FMs should be specified by target population, modality, organ system, and decision context (e.g., screening, triage, response assessment), and evaluated under conditions that reflect clinical use: heterogeneous scanners and protocols, real prevalence, local workflows, and downstream actions. Claims of general-purpose capability without decision-context grounding invite silent misuse, especially when models are repurposed beyond their validated scope. REAL-FM's paired indicators of hype and reality (Table 1) offer a practical diagnostic for identifying such misalignment before deployment.

Responsible deployment requires explicit objective function tuning, as optimizing for clinician guideline adherence (e.g., high sensitivity) often creates inverse correlations with payer cost-containment goals. Responsible deployment is not only a technical problem; it is a values problem[125]. Clinical decisions are inherently preference-sensitive, and even a highly competent system can fail if the priorities embedded in its recommendations are misaligned with patient values, undermining both trust and efficacy[126]. In parallel, modern AI systems can be steered to reflect different stakeholder objectives, because values enter through training data, tuning, and usage context; consequently, the same underlying model may produce different recommendations when optimized or prompted for clinician guideline adherence versus payer cost constraints or patient/family risk tolerance[125]. For imaging FMs, this implies that optimization choices (e.g., sensitivity vs specificity, false reassurance vs over-triage, throughput vs deliberation) are not neutral; they encode priorities. Therefore, value transparency must be explicit: who the model is designed to serve, what trade-offs it encodes, and how those choices are governed and updated over time[126].

At the modeling level, the one-model-fits-all narrative deserves caution. Imaging is heterogeneous across devices, institutions, and populations; simply aggregating large, diverse datasets does not guarantee robustness and can conceal systematic bias. Instead, we argue for a bottom-up, curriculum-style pathway: start with a single modality and anatomical domain, mature performance across clinically defined tasks, then expand (i) across modalities within a fixed organ system and (ii) across organs with a fixed modality stack. This staged approach supports not only technical generalization but also repeated re-specification of clinical objectives and acceptable error trade-offs as the scope and the value landscape change[125].

Because data-driven learning is stochastic and shaped by observational bias, imaging FMs should incorporate domain constraints wherever feasible: well-defined sub-tasks aligned with physiology and pathology, mechanistic or anatomical priors, and causal analyses that probe why predictions arise rather than only what they predict. In practice, the most defensible role for FMs is to scale high-volume perception and measurement (standardized detection, segmentation, quantification, and longitudinal tracking) while keeping integrative, value-laden decisions and trade-offs anchored in clinician–patient judgment.

Current FMs lack the symbolic reasoning required for autonomous diagnosis; therefore, near-term utility resides in hybrid architectures that couple generalist backbones with constrained, rule-based modules. As one practical instantiation, neuro-symbolic AI couples imaging FMs with ontologies, clinical rules, and logical constraints to enforce guideline-consistent reasoning, improve causal/consistency checks, and strengthen clinical trust (***Suppl Fig. 4***). These designs can produce structured evidence and uncertainty in a form that clinicians can verify and contest, rather than replacing deliberation with opaque outputs.

Operationalizing imaging FMs requires integrating causal discovery algorithms with uncertainty quantification to ensure model stability under distribution shift. Value alignment is essential because models can be tuned and deployed in ways that shift priorities across stakeholders[125]. Trustworthiness requires more than high average accuracy: calibrated uncertainty, subgroup robustness, stability under distribution shift, and transparent failure modes. Safety must be designed in, with rigorous pre-deployment stress testing and continuous post-deployment monitoring. If pursued with staged development, explicit intended use, and transparent governance of clinical trade-offs, systematically assessed through frameworks such as REAL-FM, imaging FMs can evolve from impressive pattern-recognition engines into accountable collaborators that help clinicians see more, measure more, and discover new clinically meaningful endpoints.

## Author contributions

AM, KZ, RQ, SMA and JW conceived the scope and central thesis of the Perspective. AM led the conceptual development of the manuscript, including the critical analysis of foundation models, the taxonomy of reasoning, and the discussion on causality, trustworthiness, and deployment challenges in biomedical imaging. AM, KZ, RQ, and SMA contributed to the evaluation of current foundation model paradigms, limitations, and emerging trends across imaging modalities. SF, HA, SMA, and MW provided domain expertise on clinical relevance, validation practices, and translational considerations. JW supervised the overall direction of the work and provided strategic guidance on clinical and methodological framing. All authors contributed to the writing, critical revision, and intellectual refinement of the manuscript, and approved the final version for publication.

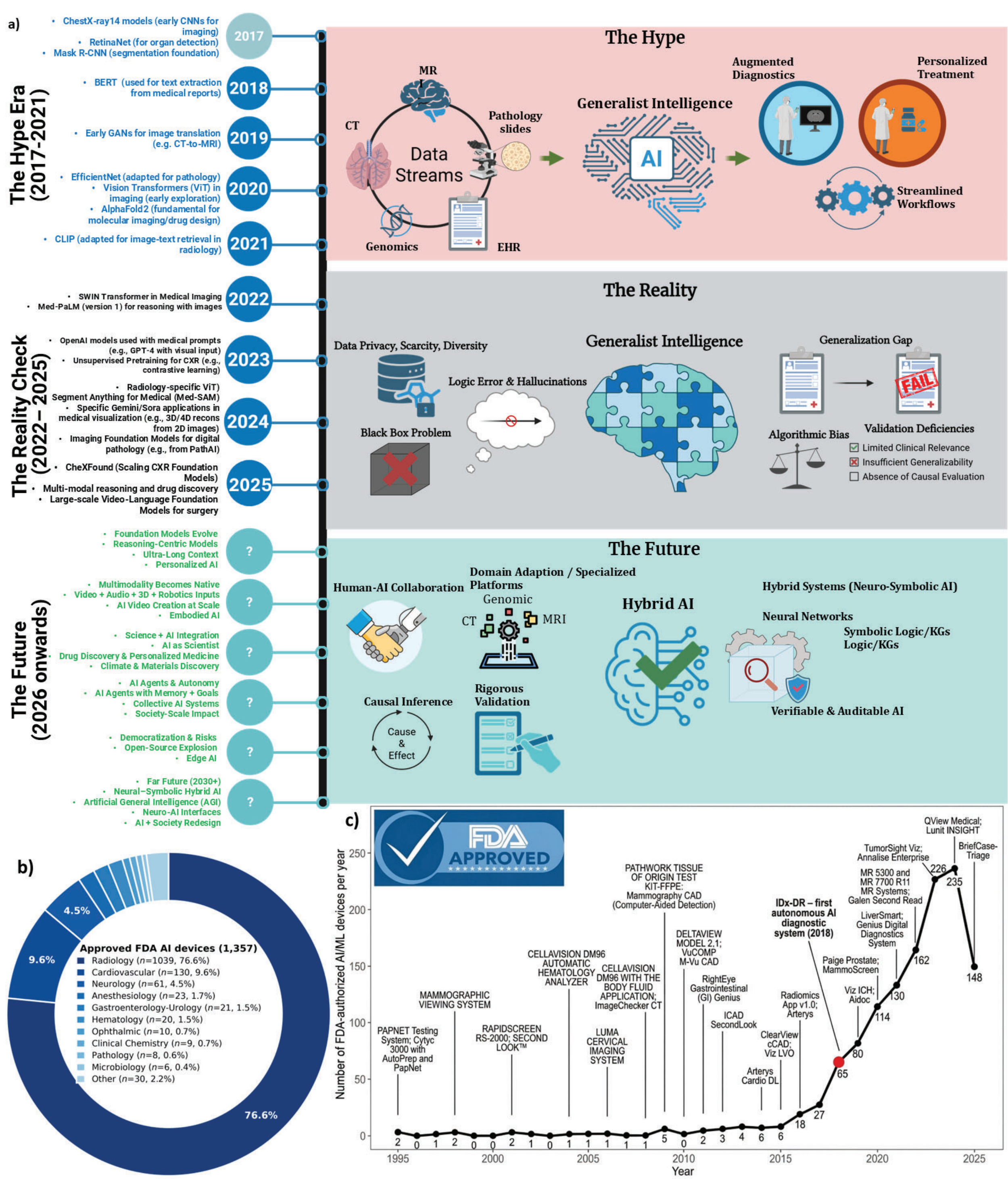

**Figure 1: Evolution of artificial intelligence in biomedical imaging, from innovation to clinical translation**. (**a**) Overview of the field across three phases: the *Hype Era* (2017–2021), the *Reality Check* period (2022–2025), and projected future directions (2026 onwards). Early

transformer-based and self-supervised advances fueled enthusiasm for general-purpose models, whereas later multimodal and reasoning-oriented systems exposed major translational barriers, including data privacy, limited generalizability, validation deficiencies, algorithmic bias and hallucinations. The future phase highlights hybrid AI, causal inference, rigorous validation, human–AI collaboration and domain-specialized biomedical FMs. (**b**) Distribution of FDA-authorized AI/ML-enabled medical devices by clinical domain. (**c**) Annual FDA authorizations of AI/ML-enabled medical devices from 1995 to 2025, with one representative example of approved systems per year. Source data for **b,c** were obtained from the U.S. Food and Drug Administration database of AI/ML-enabled medical devices; data collection cutoff, **8 January 2026**.

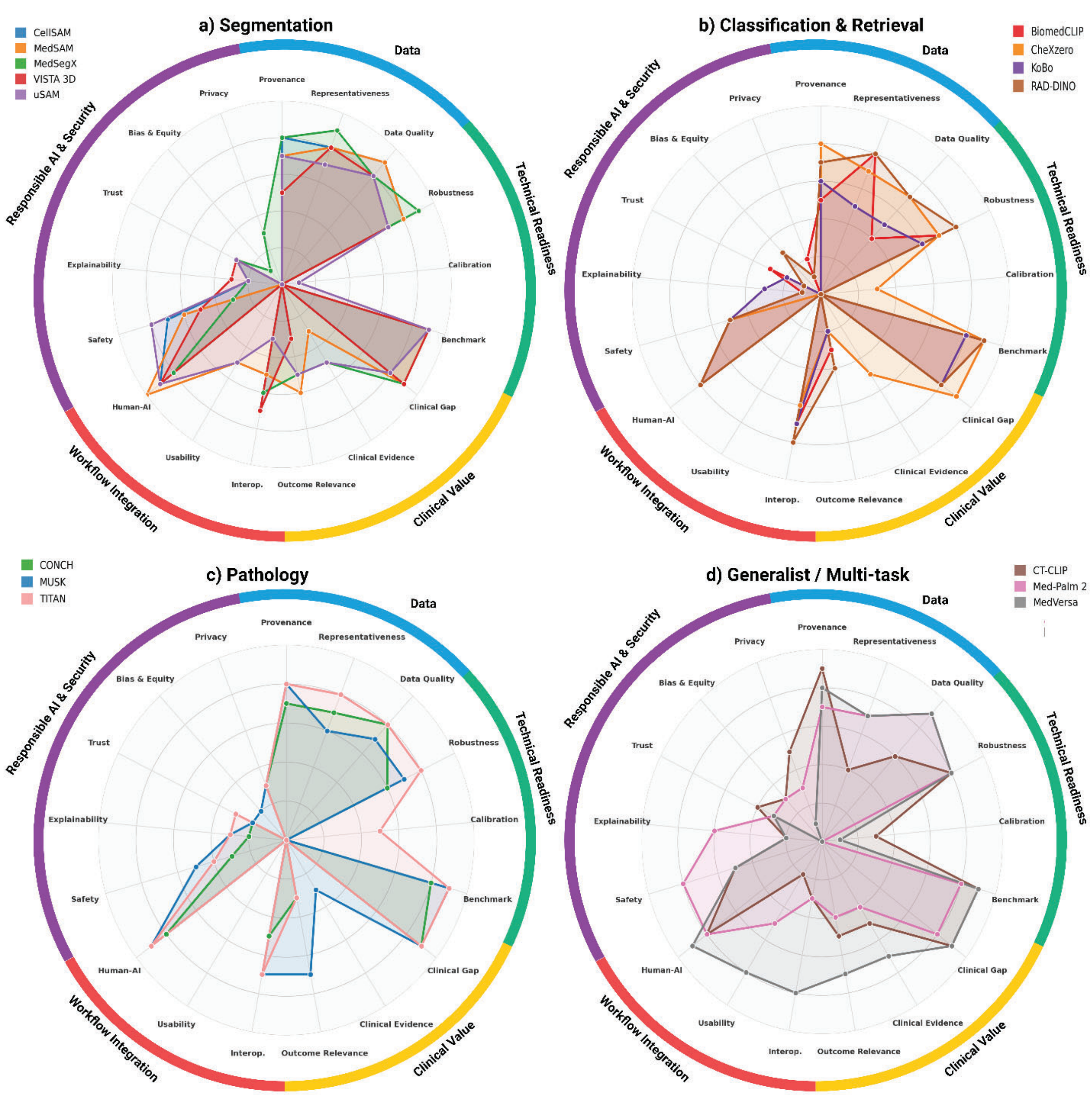

**Figure 2: Comparative evaluation of biomedical FMs across domains**. Radar plots summarize representative models in a) segmentation, b) classification and retrieval, c) pathology, and d) generalist or multi-task settings across key dimensions of translational readiness, including data, technical performance, clinical value, workflow integration, and responsible AI. Larger radial extent indicates stronger support for real-world readiness on a given criterion**.** While several models achieve broad strength in benchmark-oriented dimensions, substantial gaps remain in clinically critical areas such as calibration, usability, interoperability, privacy, accountability, and bias and equity assessment, underscoring the persistent gap between technical progress and real-world clinical translation.

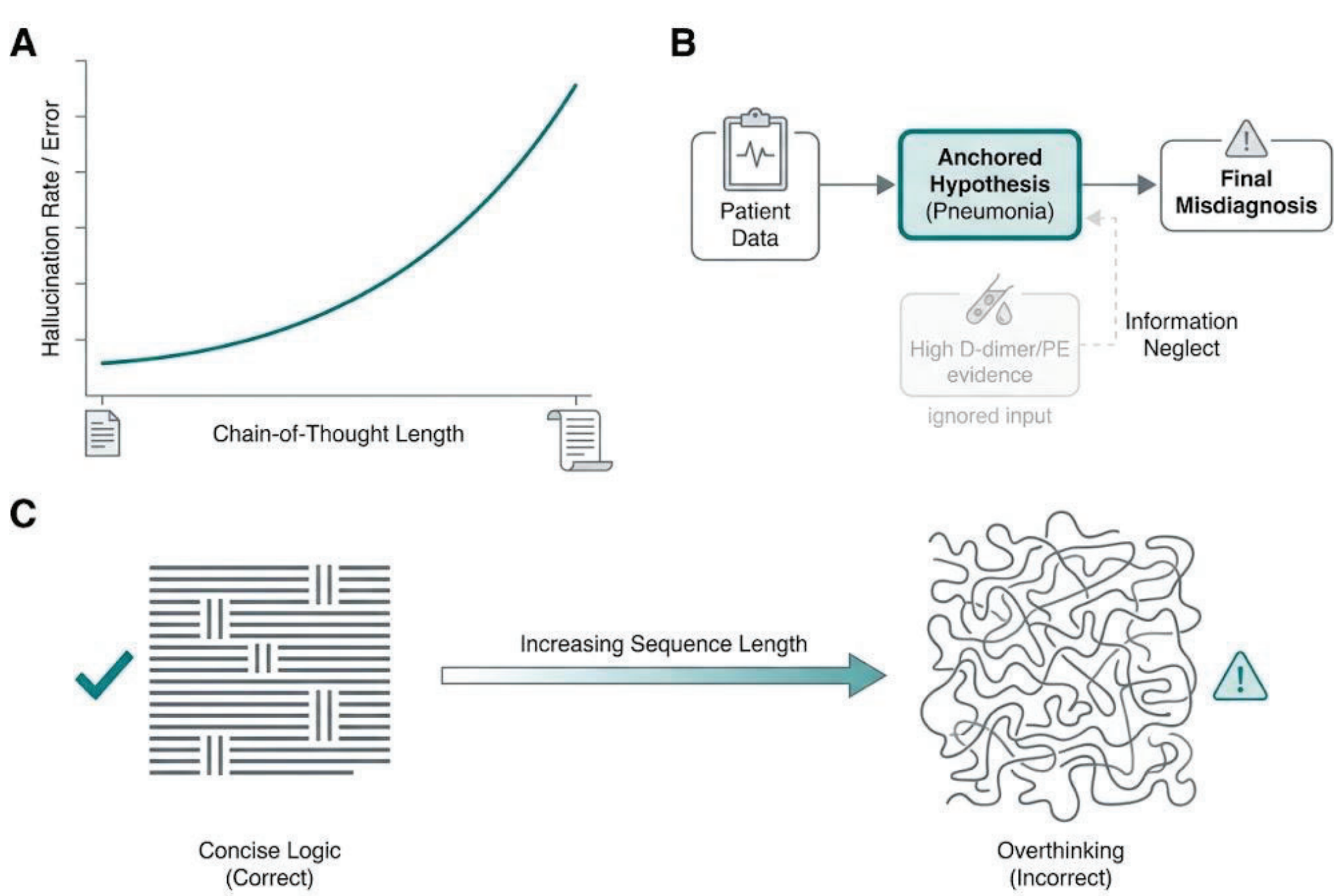

**Figure 3: Longer reasoning chains can increase error and reveal human-like cognitive biases in FMs**. (a) Conceptual relationship between reasoning length and error rate: as chains of reasoning become longer and more elaborate, empirical studies show a rising probability of incorrect answers (overthinking). (b) Example of anchoring and failure to integrate conflicting evidence: an initial hypothesis of pneumonia is formed from fever, cough, and chest opacity, and is retained as the final output despite new high-value evidence (leg swelling and elevated D-dimer) that supports an alternative diagnosis of pulmonary embolism, which is effectively ignored. (c) Schematic comparison of short versus long explanations: a concise reasoning trace yields the correct answer, whereas an extended explanation, though more verbose and seemingly

sophisticated, remains incorrect, illustrating that verbose rationalization does not imply deeper reasoning and can instead amplify an early mistaken hypothesis.

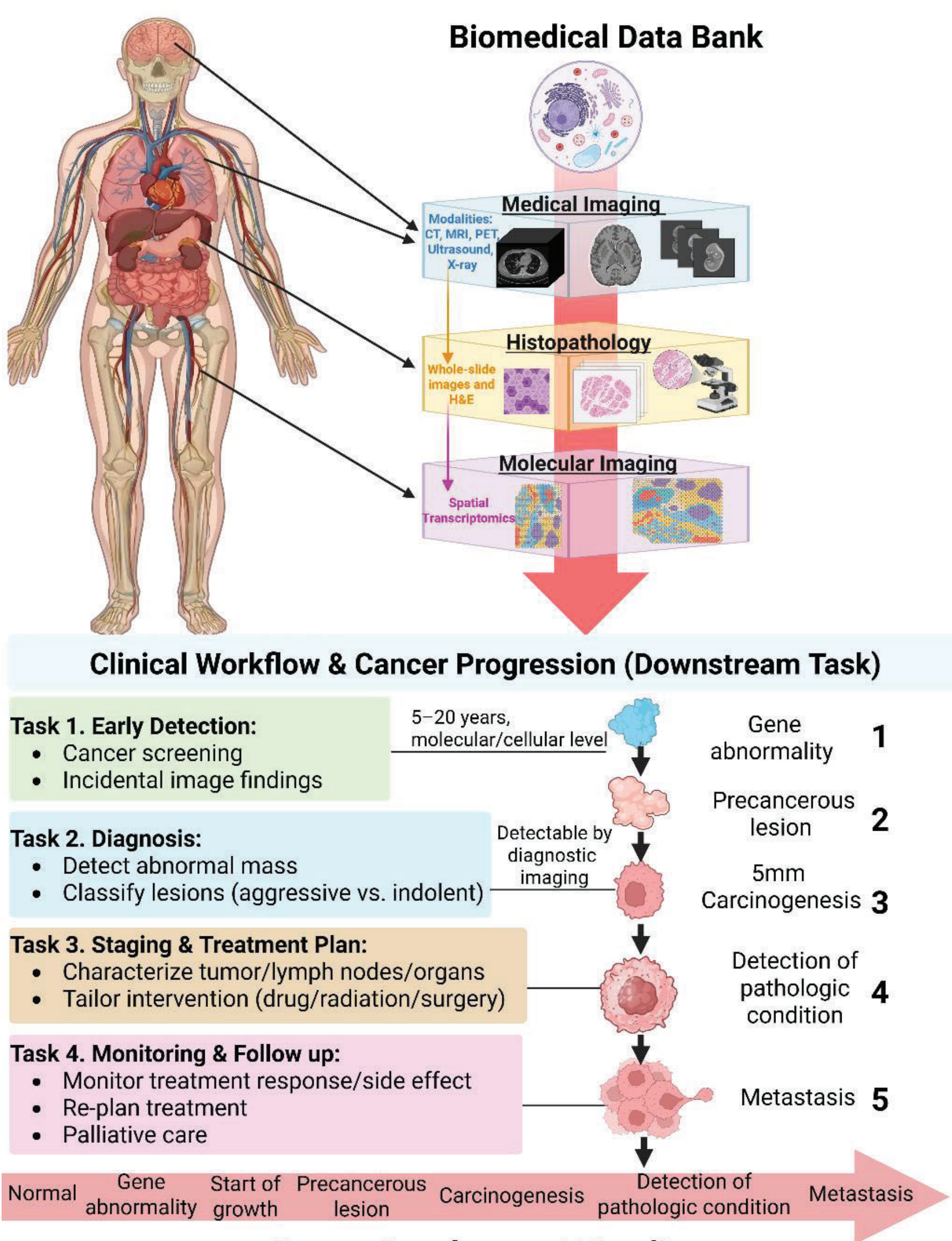

**Figure 4**: **Multimodal biomedical data bank and downstream imaging tasks across the cancer trajectory**. A central biomedical data bank from medical imaging to molecular imaging represented in the context of cancer applications. The lower panel maps four imaging-driven tasks

(early detection, diagnosis, staging and treatment planning, and monitoring/follow-up) to corresponding stages of tumor progression along the cancer development timeline.

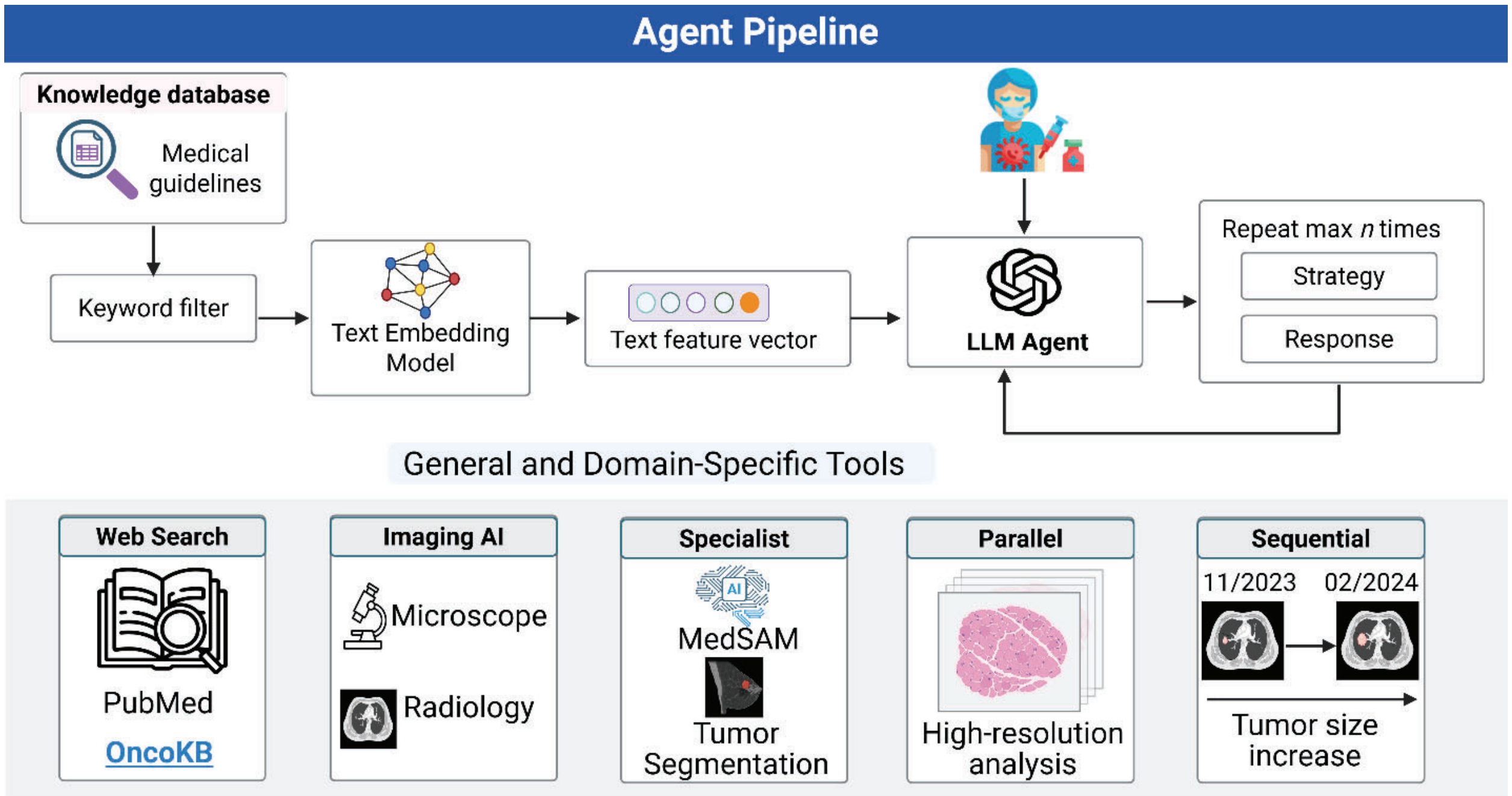

**Figure 5**: **Specialist agent pipeline in oncology**. Medical guidelines are converted into text embeddings and combined with an individual patient case as input to an LLM-based agent. The agent can call external tools (e.g., PubMed/OncoKB search, imaging AI, and tumor-segmentation models) to analyze images in parallel (e.g., pathology slides) and sequentially over time (e.g., serial CT scans).

**Table 1: Evaluation Framework REAL-FM for Biomedical Foundation Models: Hype vs. Reality.** This framework categorizes criteria across five axes to distinguish superficial performance from clinical readiness. Representative data is defined by multi-site, multi-vendor coverage with quantified demographics and documented gaps. Clinically meaningful endpoints prioritize real-world outcomes (e.g., changes in diagnosis or morbidity) over surrogate metrics like AUC or Dice. While some axes share conceptual boundaries, we distinguish between technical robustness, which assesses general performance stability under distribution shift, and bias/equity, which evaluates whether that performance remains equitable across specific demographic subgroups.

| Axis | Criterion | Indicators of Hype | Indicators of Reality |
|---|---|---|---|
| **Data** | *Provenance/ Transparency* | Data described only as large or diverse with no details on sites, scanners, time period, or labelling process. | Detailed data sheets document institutions, demographics, scanners, time frame, inclusion/exclusion criteria, and labelling protocol. |
| | *Representativeness* | Single-center or single-vendor data presented as broadly representative; no analysis of demographics, disease mix, or domain shift. | Multi-center, vendors, and populations, with quantified coverage of relevant demographics, disease spectrum, and imaging protocols, and explicit discussion of residual gaps and domain shift. |
| | *Data/Label Quality* | Focus on scale; labels are weak/noisy (e.g., report-derived) with no QC or inter-rater statistics. | Expert or adjudicated labels; inter-rater agreement and label noise are quantified, and quality-control procedures are described. |
| **Technical Readiness** | *Model Robustness* | Performance reported only on a single internal test split; no external or out-of-distribution evaluation, no failure analysis. | Robustness demonstrated on external and distribution-shifted cohorts, with stress tests and qualitative analysis of failure modes. |
| | *Calibration/ Uncertainty* | Deterministic outputs: only AUC/accuracy reported; no calibration metrics, uncertainty estimates, or abstention behavior. | Probabilities are calibrated and evaluated, uncertainty estimates are explicitly incorporated into thresholds, triage rules, or “I don’t know” policies that are meaningful for clinical use. |
| | *Benchmark Integrity/ Validation* | State-of-the-art weak baselines, or potential test leakage; clinically irrelevant operating points. | Uses public or preregistered splits with strong baselines, temporally separated or prospective validation, and evaluation at clinically relevant operating points. |
| **Clinical Value** | *Clinical Gap* | The task is chosen for technical convenience and presented as important without a clearly articulated unmet clinical need or workflow bottleneck. | A specific clinical gap or workflow bottleneck is articulated, and the model’s intended role in the care pathway is clearly specified. |
| | *Clinical Evidence* | Claims of utility rest solely on retrospective test metrics; no comparison with standard-of-care readers or real-world use. | Multi-site retrospective validation plus reader-in-the-loop and/or prospective studies comparing performance to standard-of-care practice. |
| | *Decision-Making Impact/ Outcome Relevance* | Evaluation stops at surrogate metrics (e.g., AUC, F1); no assessment of changes in decisions or outcomes. | Impact is assessed using clinically meaningful endpoints such as detection rates, time-to-diagnosis, downstream utilization, and/or patient outcomes, with discussion of cost-effectiveness where feasible. |
| **Workflow Integration** | *Interoperability* | The tool is a standalone application or widget that disrupts the existing clinical workflow. | Seamless, standards-based (e.g., DICOM, HL7) integration with existing PACS, RIS, and EHR systems. |
| | *Usability* | No evidence of testing with actual clinicians in a real-world setting. | Formal usability studies with target end-users have been conducted and demonstrate efficiency and satisfaction. |
| | *Human-AIInteraction* | The model is positioned as an autonomous decision-maker or replacement for the clinician. | The system is designed as an augmentation tool with clear mechanisms for human oversight, feedback, and final judgment. |
| **Responsible AI & Security** | *Safety & Hallucination* | Generative outputs (reports, recommendations, rationales) are deployed without quantifying hallucination rates or clinical severity; safety is delegated to disclaimers and manual review. | Hallucinations are measured on representative tasks; mitigation strategies (e.g. retrieval augmentation, constrained generation) and mandatory human verification for high-stakes use are in place. |
| | *Explainability And Reasoning* | Post-hoc saliency maps or anecdotal examples are presented as explanations without testing fidelity, stability, or usefulness for clinicians. | Explanation methods are evaluated for faithfulness and robustness, and their impact on clinician understanding, trust, and error detection is studied. |
| | *Trust & Accountability* | Governance, logging, and incident response are not described; responsibility for failures is implicitly shifted to individual clinicians or vague AI assistance language. | Clear accountability is allocated across developers, vendors, and institutions; |
| | *Bias & Equity* | Fairness is mentioned rhetorically, but performance is not stratified by sex, age, ethnicity, site, or socioeconomic status. | Performance is reported across key subgroups and sites; disparities are analyzed, and mitigation strategies are discussed. |
| | *Data Privacy, Cybersecurity* | Generic statement (e.g. data are anonymized and stored securely), with no detail on de-identification, encryption, or threat modelling. | Concrete description of de-identification, encryption, access control, and security testing; consideration of federated/on-prem deployment and regulatory compliance (e.g., HIPAA/GDPR). |

**Table 2:** A Taxonomy of Reasoning in Biomedical Foundation Models.

| Reasoning Paradigm | Core Capability | Key Methodologies | Exemplar Frameworks/Models | Hype | Reality Check |
|---|---|---|---|---|---|
| **Implicit & Sequential** | Emulating the step-by-step diagnostic workflow and generating explanatory rationales. | Chain-of-Thought (CoT) Prompting, Hierarchical Verification, and Visual Grounding. | Med-PaLM 2[57], MedCoT[58] , CoVT[59]. | AI thinks like a doctor. | AI mimics the narrative structure of a doctor's report, a linguistic feat that can be plausible but factually incorrect. |
| **Spatial** | Understanding and interpreting anatomical structures, their relationships, geometries, and pathological changes in 2D and 3D space. | Large-scale Segmentation, Adaptive Convolutions, Mixture-of-Experts, Geometric Deep Learning, Spatial Tokens, Reasoning Tokens | SAM-Med2D[22], SPAD-Nets[71], MoME[72], PointNet-based models[70]. | A single generalist model can understand complex, high-dimensional spatial context medical images. | The heterogeneity of medical data necessitates domain-adapted, architecturally flexible, spatially grounded, or federated specialist models. |
| **Explicit & Symbolic** | Integrating verifiable domain knowledge to improve robustness, interpretability, and evaluation. | Knowledge Graph (KG) Integration, Neuro-Symbolic Logic, Causal Inference. | KoBo[91], ReXKG[93], CURE AI[127], NSAI Frameworks. | AI develops intelligence through scale alone. | Trustworthy medical AI requires a return to symbolic systems to inject and verify logical, causal, and factual knowledge. |

**Table 3:** State-of-the-Art Foundation Models and their Evaluated Reasoning Capabilities.

| Model/ Framework | Primary Modalities | Claimed Reasoning Advance | Key Evaluation Results (The Reality Check) |
|---|---|---|---|
| Med-PaLM 2[57] | Text, Vision+Text | Ensemble refinement and chain-of-retrieval for advanced medical question answering. | Achieves 86.5% on MedQA (USMLE). Clinicians preferred its answers over generalist physicians on the reasoning axis, but specialist answers were still superior overall. Safety is rated as on par with physicians, not better. |
| MedCoT[58] | Vision+Text | Hierarchical expert verification to improve VQA accuracy and interpretability by mimicking multi-physician review. | Outperforms single-expert models on Med-VQA datasets. However, it relies on a multi-stage, engineered process. |
| CoVT[59] | Vision+Text | Chain-of-Visual-Thought framework for interpretable diagnosis by grounding textual reasoning in visual prompts. | Demonstrates improved performance over baselines on the custom CoVT-CXR dataset. The approach requires a massive, fine-grained annotation effort, showing that deep interpretability is not a free lunch. |
| SAM-Med2D[22] | Vision | Large-scale adaptation of the SAM for universal 2D medical image segmentation. | Fine-tuned on 4.6 million medical images. While it improves over the generalist SAM, its performance on specific tasks is often still inferior to domain-specific models, challenging the one model fits all hype. |
| KoBo[91] | Vision+Text | Knowledge-boosting pre-training that integrates a clinical knowledge graph to mitigate spurious correlations. | Outperforms baseline VLP methods on eight downstream tasks. Demonstrates the necessity of injecting explicit knowledge to overcome limitations of purely data-driven learning. |
| MedVersa[96] | Vision+Text | A generalist foundation model for radiology that achieves competitive performance with specialized solutions. | Radiologist evaluations showed that reports matched or exceeded human reports in 71% of cases. However, deeper analysis shows significant gaps in capturing complex relationships between entities compared to human experts. |
| Med-CLIP[97] | Vision+Text | Contrastive pre-training from unpaired medical images and reports, using a knowledge-guided semantic matching loss to reduce false negatives and boost data efficiency. | Outperformed all previous methods on zero-shot prediction, supervised classification, and image-text retrieval, and won over SOTA using 10% of the data. Therefore, excellent data efficiency and transferability, but zero-shot clinical use is limited by prompt sensitivity, and further gains require more pretraining data. |
| RAD-DINO[128] | Vision | Demonstrates that high-quality and general-purpose medical image representations can be learned without any text supervision. | RAD-DINO, trained on images only, achieves or exceeds SOTA across classification, segmentation, and vision–language tasks, proving text supervision is not necessary for high-quality medical image encoders. |
| VISTA 3D[129] | Vision+Text | Unifies a 3D FM for automatic and interactive segmentation, enabling efficient human correction, zero shot handling of unseen structures, and a full annotation workflow that reduces manual effort. | Achieves SOTA 3D automatic segmentation across 127 classes and strong interactive performance, with its 3D supervoxel design boosting zero-shot accuracy by ~50% over baselines. |
| PubMedCLIP[130] | Vision+Text | Produces organ- and modality-aware image representations for more precise, context sensitive VQA and retrieval in MedVQA pipelines, especially for complex multi-organ and rare scenarios, though it lacks explicit reasoning or KG integration | Improves medical VQA benchmarks by up to 6% and delivers ~40% relative gains in image–text retrieval versus CLIP. Still, the text encoder trails specialist QA models and overall interpretability, and robustness remains limited under real-world domain shift. |
| KAD[131] | Vision+Text +Knowledge Graph | Integrates structured medical knowledge (UMLS, RadGraph) with image report analysis to support zero-shot disease queries, interpretable diagnostic attention, and knowledge guided, multi-step reasoning aligned with clinical workflows. | Achieves SOTA zero-shot diagnostic accuracy on multiple external chest X-ray datasets and outperforms prior models in few-shot settings (down to 1% labeled data), but depends on a validation set for thresholding, cannot detect diseases absent from the knowledge base, and does not produce segmentation masks. |
| CheXzero[33] | Vision+Text | Self-supervised contrastive pre-training on paired chest X-rays and radiology reports for zero-shot pathology classification without explicit labels. | Matches radiologist-level performance for five CheXpert pathologies, with mean AUC only 0.042 below fully supervised SOTA and outperforming all prior self/semi-supervised methods using 0 labeled reports; robustness still depends on report quality and prompt phrasing. |
| BiomedCLIP[25] | Vision+Text | Large-scale domain-adaptive VLP. | Achieves SOTA cross modal retrieval and strong zero-shot classification/VQA, surpassing PubMedCLIP and MedCLIP on several benchmarks, but its generalist training can underperform for niche specialties or rare diseases and still requires careful prompt design. |
| CONCH[80] | Vision+Text | Joint WSI–caption representation with heatmaps. | Reaches SOTA zero- and few-shot accuracy on 14 slide- and ROI-level WSI benchmarks, though performance on rare or very fine-grained subclasses remains challenging. |

## Supplementary Materials

### A. Figures

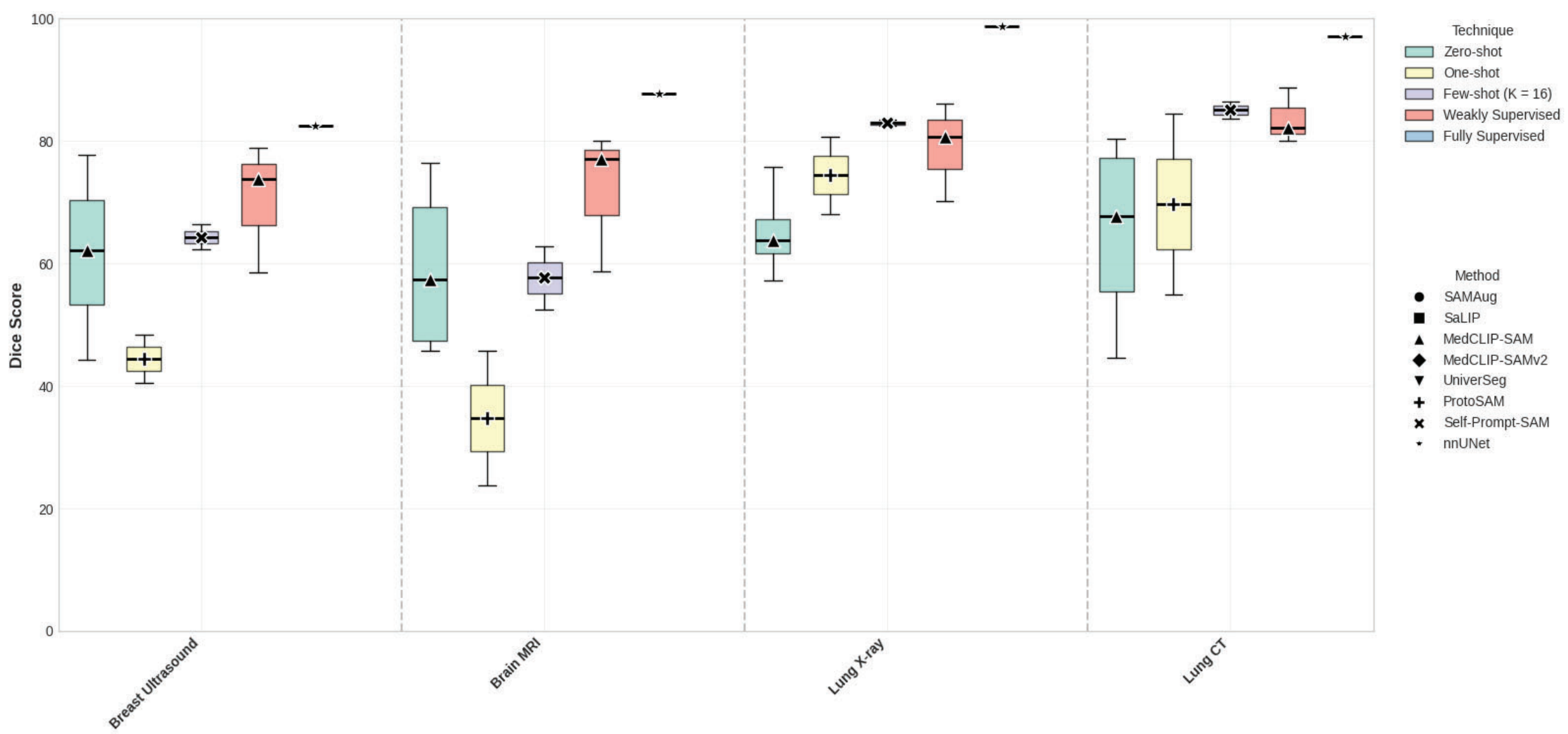

**Suppl Fig. 1:** Benchmarking segmentation performance across various biomedical FMs adapted from the literature. Dice scores are reported for four imaging modalities: breast ultrasound, brain MRI, lung X-ray, and lung CT. The color indicates the model training technique, and the marker indicates AI models.

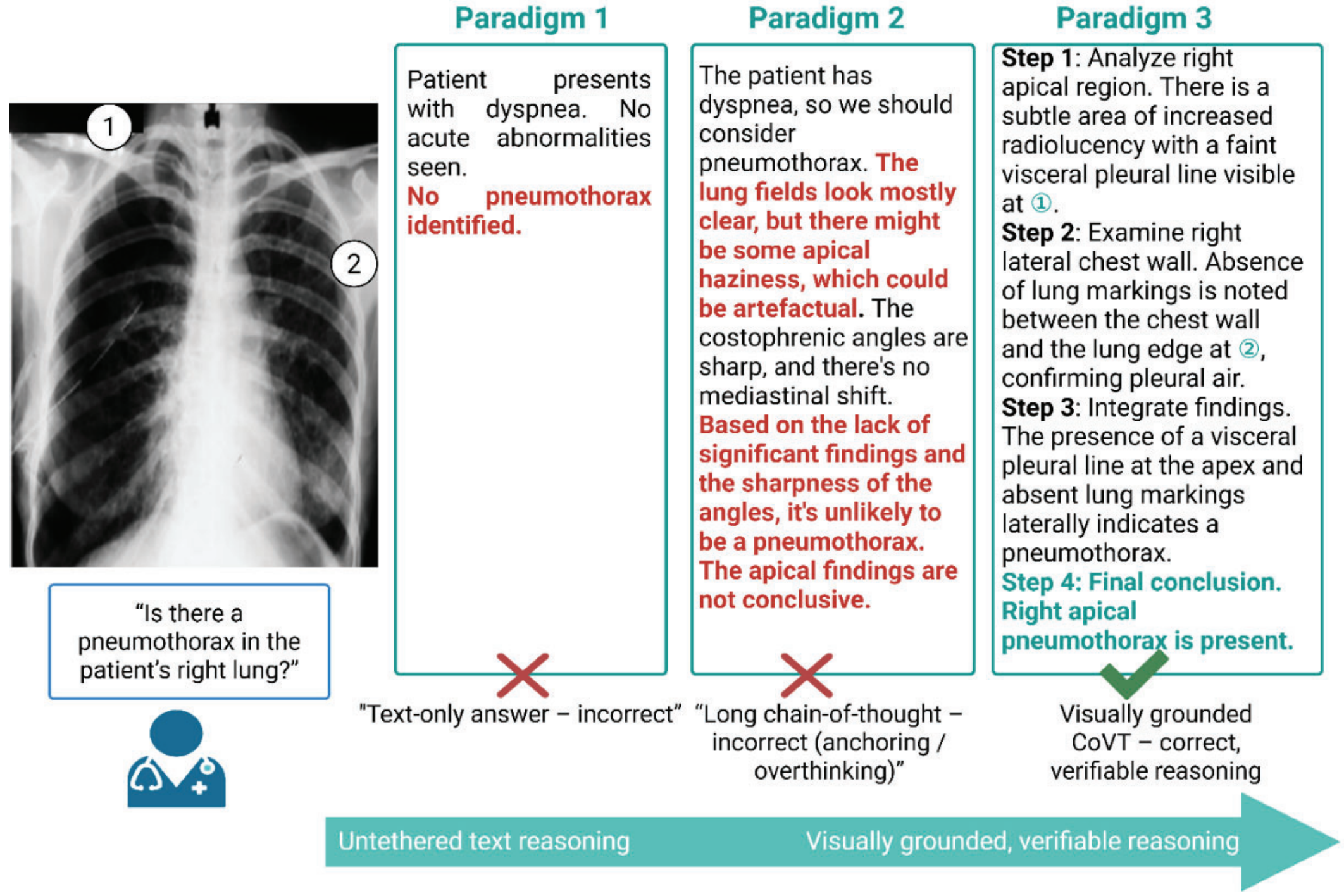

**Suppl Fig. 2:** Comparing text-only reasoning and Chain-of-Visual-Thought (CoVT) for pneumothorax detection. Given a chest radiograph and the query "Is there a pneumothorax in the patient's right lung?", a short text-only report (Paradigm 1) and a long, unconstrained chain-of-thought explanation (Paradigm 2) both arrive at the wrong conclusion due to failing to attend to key apical and lateral visual cues. In contrast, Paradigm 3 (CoVT) decomposes the decision into explicit, stepwise statements, each grounded in specific image regions, leading to the correct diagnosis of a right apical pneumothorax. The bottom bar illustrates the conceptual shift from untethered text reasoning to visually grounded, verifiable reasoning.

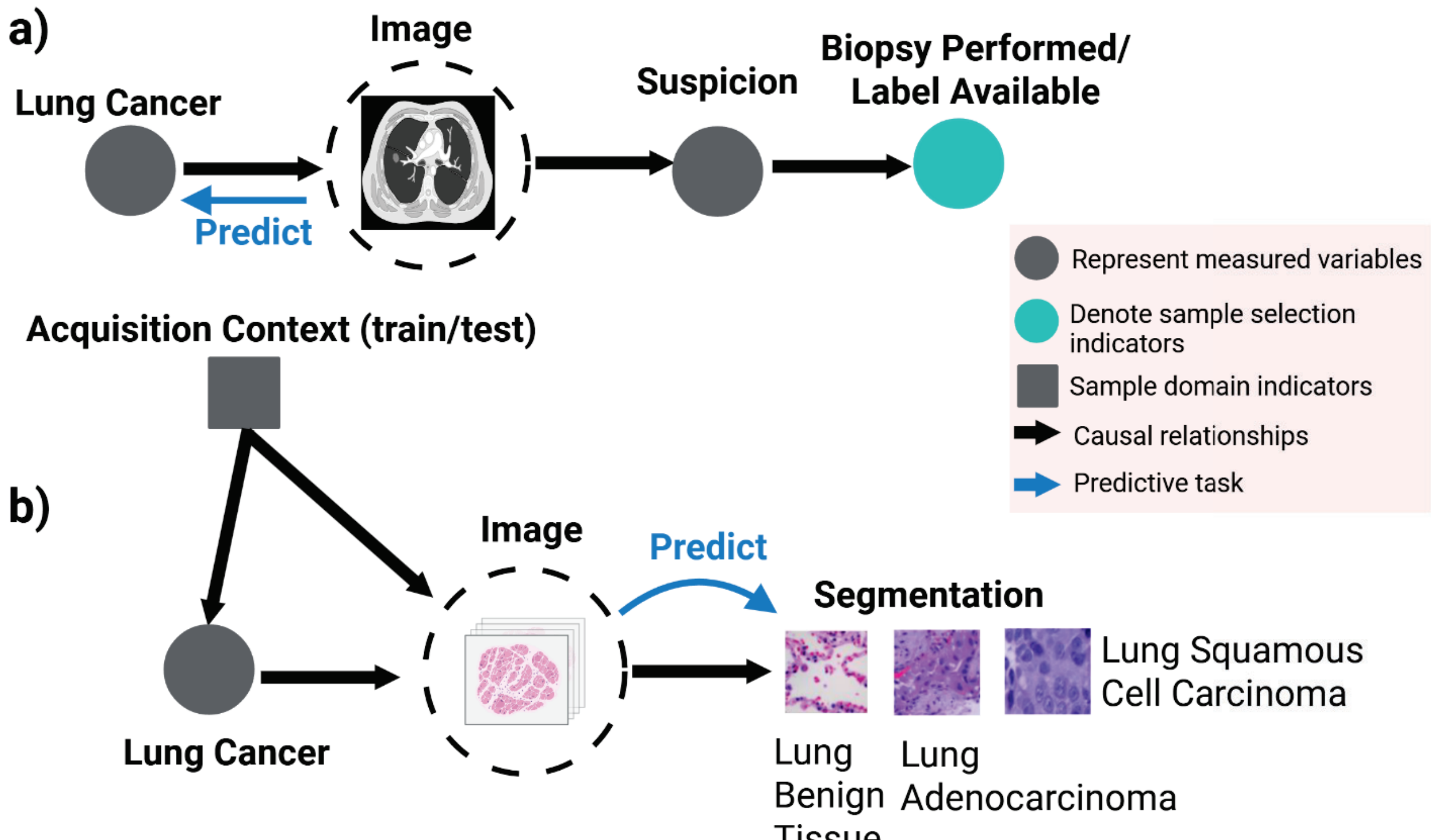

**Suppl Fig. 3**: **Causality in biomedical imaging.** (a) Lung cancer prediction (blue arrow) as a statistical inference task where the underlying disease causally drives clinical suspicion, which in turn triggers biopsy and label generation (black arrows), revealing how selection bias can confound model training. (b) Lung tumor segmentation where the acquisition context serves as a domain indicator affecting both disease manifestation and image generation. Filled circles represent measured variables, cyan circles denote sample selection indicators, and gray squares represent sample domain indicators. Blue arrows indicate the direction of the predictive task, while black arrows represent causal relationships.

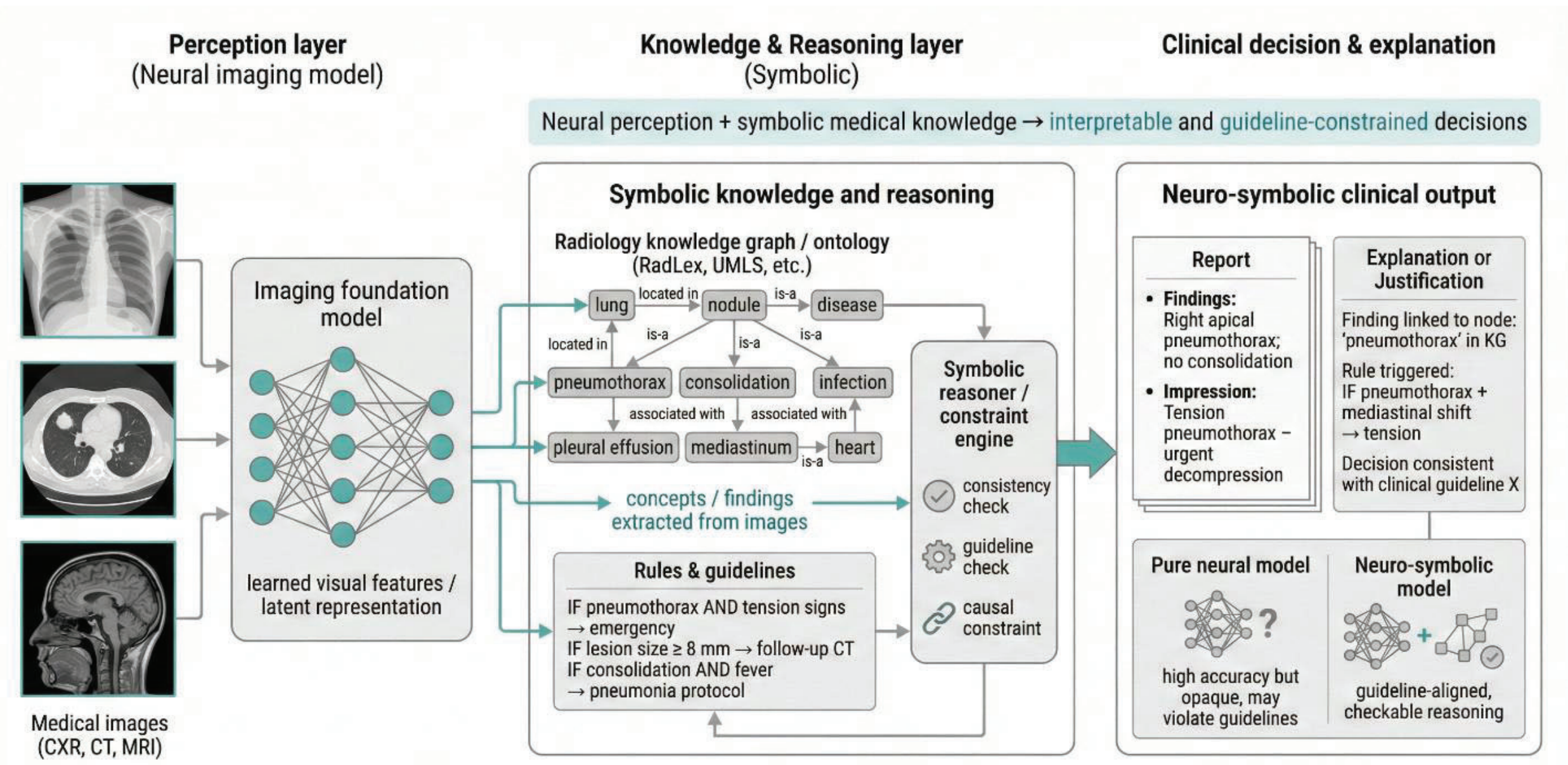

**Suppl Fig. 4:** Neuro-symbolic AI combines imaging foundation models with ontologies, rules, and logical reasoning to improve robustness, causal consistency, and clinical trust.

## B. Supplementary Information

**Supplementary Note 1**. To improve the auditability of radiology report generation, CoVT-CXR was developed using a carefully annotated chest X-ray dataset in which reasoning steps in the report are aligned with specific image regions. This design encourages the model to learn a direct association between semantic claims, such as the presence of cardiomegaly, and the image evidence supporting those claims. As a result, the reasoning process becomes not only more interpretable but also visually auditable.

**Supplementary Note 2**. The Radiology Knowledge Graph system (ReXKG) extends evaluation from the surface text level to the knowledge level.[1] It works by first extracting structured information (entities and their relationships) from an AI-generated report to construct a KG. This generated KG is then compared against a ground-truth KG derived from a human expert's report. Using novel metrics that assess the similarity of nodes (entities), edges (relationships), and subgraphs, ReXKG can evaluate whether the AI model has correctly understood not just the key findings but also their clinical context and interrelations, a much deeper and more meaningful assessment of its reasoning capabilities.[1]

**Supplementary Note 3**. From tool to teammate: agentic AI in clinical practice. The field is moving from AI as a passive tool that responds to isolated prompts toward AI as an autonomous agent capable of completing complex clinical workflows. This transition reflects the difference between a system that provides a single prediction and one that can operate as a clinical teammate[2].

Traditional AI systems are largely reactive: a clinician provides an input, such as an image, and the system returns a segmentation, classification or probability score. Human users remain responsible for supplying context, interpreting outputs and deciding the next steps. An AI agent, by contrast, is a proactive, goal-oriented system. Given a high-level objective, such as developing a treatment plan for this patient, an agent can perceive the necessary information, reason about a course of action, formulate a plan, and execute it by interacting with its environment, including other AI models, databases, and software tools. This functionality is typically enabled by an agentic framework composed of several core modules: a perception module for data ingestion, a memory module for maintaining context, a reasoning module for planning, and a tool-use module for executing actions.

**Supplementary Note 4**. Neuro-symbolic AI (NSAI) represents an emerging strategy for combining data-driven learning with explicit knowledge-based reasoning[3]. This hybrid approach directly addresses the black box problem by combining the perceptual power of neural networks with the logical rigor of symbolic AI. In a typical NSAI architecture for medical imaging, a neural network component would act as a powerful perception engine, analyzing the raw image pixels to identify objects and features. This structured information is then passed to a symbolic reasoning engine, which operates on formal logic, ontologies, and rule-bases to validate the findings, draw inferences, and construct a transparent, auditable explanation for the final diagnosis. This paradigm represents a full-circle evolution of AI, moving away from the brittle, rule-based expert systems of the past, through the powerful but opaque era of deep learning, and now towards a hybrid future that promises to combine the best of both worlds. This trajectory suggests that for FMs to become truly trustworthy partners in medicine, they cannot rely on emergent intelligence from scale alone; they must be explicitly endowed with the ability to reason with provable logic and verifiable knowledge.